\documentclass[12pt,preprint]{aastex}

\newcommand\ie{{\it i.e.,}\ }
\newcommand\etal{{\it et al.}}
\newcommand\psr{PSR~B0656+14}
\newcommand\us{\ensuremath{\mu\rm{s}}}

\newcommand\hst{{\it HST}}

\shorttitle{Photopolarimetry of \psr}
\shortauthors{Kern, Martin, \& Halpern}

\begin{document}

\title{Optical Pulse-Phased Photopolarimetry of \psr}


\author{B. Kern, C. Martin, B. Mazin}
\affil{Division of Physics, Mathematics, and Astronomy, MS 405-47, 
  California Institute of Technology,
	Pasadena, CA 91125}
\email{bdk@srl.caltech.edu}

\and

\author{J. P. Halpern}
\affil{Columbia Astrophysics Laboratory, Columbia University, 
  550 West 120th Street, New York, NY 10027}

\begin{abstract}
We have observed the optical pulse profile of PSR~B0656+14 in 10 phase bins 
at a high
signal-to-noise ratio, and have measured the linear polarization
profile over 30\% of the pulsar period with some significance.  
The pulse profile is double-peaked, with a bridge of emission between
the two peaks, similar
to gamma-ray profiles observed in other pulsars.  
There is no detectable unpulsed flux, to a 1-$\sigma$ limit of 16\% of
the pulse-averaged flux.
The emission in the bridge is highly ($\sim$ 100\%) polarized, with a position
angle sweep in excellent agreement 
with the prediction of the Rotating Vector Model
as determined from radio polarization observations.
We are able to account for the gross features of the optical
light curve (\ie the phase separation of the peaks) using both
polar cap and outer gap models.
Using the polar cap model, we are also able to estimate the
height of the optical emission regions.
\end{abstract}



\section{Introduction}

Radio pulsars, as their name implies, are well studied at radio
wavelengths.
High time-resolution pulse profiles and polarization profiles have
allowed determinations of the relative orientations of the rotation
axis, magnetic axis, and observer line-of-sight (LOS), based on
variations of the Rotating Vector Model (RVM), originally proposed
for the Vela pulsar by \citet{radhakrishnan_c1969}.
It is commonly believed that the radio emission is generated
near the surface of the neutron star \citep{kijak_g1997}, 
which simplifies the 
interpretation of the radio data.
Studies by \citet{lyne_m1988}, \citet{rankin1990}, and
\citet{everett_w2001} use different methods
and assumptions to determine these angles, with a rough
agreement established between the methods.

High-energy (infrared to gamma-ray) magnetospheric 
emission from isolated, rotation-powered radio
pulsars presents a complex theoretical picture.
Two categories of theories have dominated the explanation of 
the location of the emission region for high-energy photons.
Polar cap models \citep{daugherty_h1994,daugherty_h1996a}
claim that the emission region is close to the
neutron star surface, near the emission region for the observed
radio waves.
Outer gap models \citep{romani_y1995,romani1996}
place the high-energy emission regions farther out in the
magnetosphere, at significant fractions of the light-cylinder radius.
Part of the problem in distinguishing between these two models
lies in the small number of pulsars observed at high energies.
Pulsed gamma rays have been observed in seven radio pulsars
(counting Geminga as a radio pulsar for this argument),
with possible detections in another three \citep{thompson2001}.
Many more pulsars have been observed in X-rays \citep{becker_t1997}, 
with inconclusive impact on the question of emission model, possibly 
because of a cyclotron resonance ``blanket'' confining 
the X-rays \citep{wang_rhz1998}.
Optical pulsations have been convincingly observed in only 
three radio pulsars, namely the Crab \citep{cocke_dt1969}, 
Vela \citep{wallace_1977}, and PSR~B0540-69 \citep{middleditch_p1985},
with marginal detections in PSR~B0656+14 \citep{shearer_1997} and 
Geminga \citep{shearer_1998}.
Optical photopolarimetric measurements hold the best promise to
constrain the models for high-energy emission regions, until
such time that X-ray and gamma-ray polarimetry is available.

\psr\ is a middle-aged pulsar, with period 0.385 s and
characteristic age 1.1$\times10^5$ 
years, considered a ``cooling neutron star'' because its
soft X-ray emission is believed to come from the
neutron star surface \citep{becker_t1997}.
There is a tentative detection of a 
gamma-ray pulse \citep{ramanamurthy_fkst1996},
as well as a claim of optical pulsations \citep{shearer_1997}.
Optically, \psr\ is the second-brightest radio pulsar in the
Northern sky ($V$ = 25), after the Crab puslar.

\section{Observations}

We observed \psr\ on 2000 December 20--21 at the Palomar 5 m
telescope, with a 
Phase-Binning CCD Camera.
Over the two nights, we obtained 38,000 s of integration time, with
an average FWHM of 1.3 arcsec under non-photometric conditions.
The phase-binning CCD used is a 512$\times$512-pixel back-illuminated
CCD, masked by a 100$\times$26-pixel (50$\times$13 arcsec) slit.
Intensity is binned on-chip into 10 phase intervals using a 
periodic frame transfer of the CCD.
Between the slit and the CCD, an achromatic half-wave plate and two
broadband
polarizing beamsplitters form an imaging two-channel polarimeter,
measuring two orthogonal linear polarizations simultaneously.
The two polarized images are arranged side-by-side on the
CCD in two columns.
The half-wave plate is rotated between exposures to obtain
measurements of the linear Stokes parameters (we do not
measure $V$, the circular polarization).
All observations were through a Schott BG38 colored-glass filter, 
which together with the 
quantum efficiency of our CCD, defines a bandpass of approximately 400--600 nm.

Every 1/10 of the period of \psr\ (known {\it a priori}),
the accumulated images (both polarizations) 
are transferred away from the slit onto a
region of the CCD used for storage, but no pixels are read out of the
CCD.
At the end of 10 transfers (\ie 1 period of \psr), which alternate
between shifting
the accumulated charge both up and down on the CCD, the images accumulated
during the first phase interval are shifted back into the CCD region 
illuminated
by the slit, and the images are re-illuminated.
At any time, there are 20 accumulated images residing on the CCD
(2 columns of linearly polarized images, 10 rows of
phase-binned images), two
of which are being illuminated.

This illumination pattern is repeated for 120~s (311 cycles of \psr), at
which point the shutter is closed and the entire array is read out.
Timing of the exposures
is coordinated using a GPS receiver and time-code generator, which 
delivers 
start and stop signals
accurate to 1~\us.
An ephemeris for \psr\ from A. Lyne
(2000, private communication) was used, 
with barycentric $f = 2.5980738005954 \ \rm{s}^{-1}$, 
$\dot{f} = -3.71214\times 10^{-13} \ \rm{s}^{-2}$, 
and $\ddot{f} = 8.33\times 10^{-25} \ \rm{s}^{-3}$ at 2451687.5 JD TDB, 
and a radio pulse geocentric time-of-arrival 2451687.500000997 JD UTC.
The JPL DE200 solar system ephemeris was used to compute the 
barycenter-geocenter and geocenter-topocenter corrections.
On each night, we observed the Crab Pulsar to verify our timing
and polarization optics.

\section{Data Analysis}

\begin{figure}
\epsscale{1.0}
\plotone{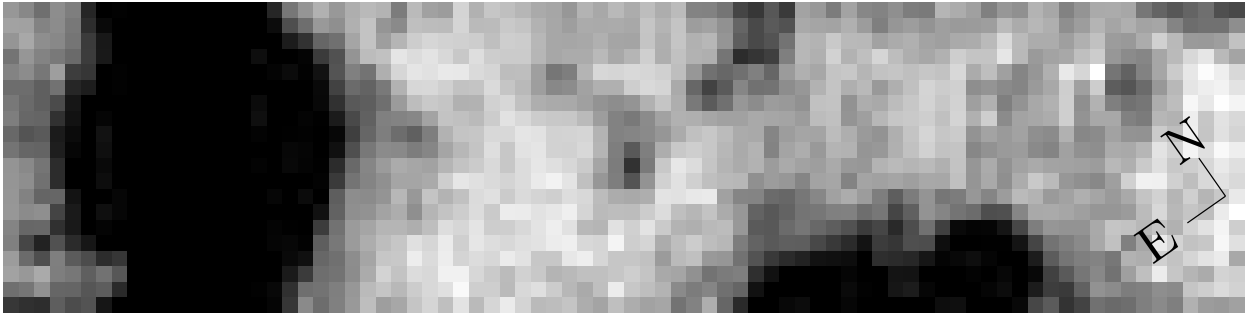}
\caption{Average intensity image, eliminating all phase and polarization
information.
\psr\ is in the exact center of the image.
Note the extended object which overlaps the pulsar 
(above the pulsar in this plot).
The size of this image is 40$\times$10 arcsec (0.5 arcsec/pixel).
\label{imgtot}}
\end{figure}

\begin{figure}
\epsscale{1.0}
\plotone{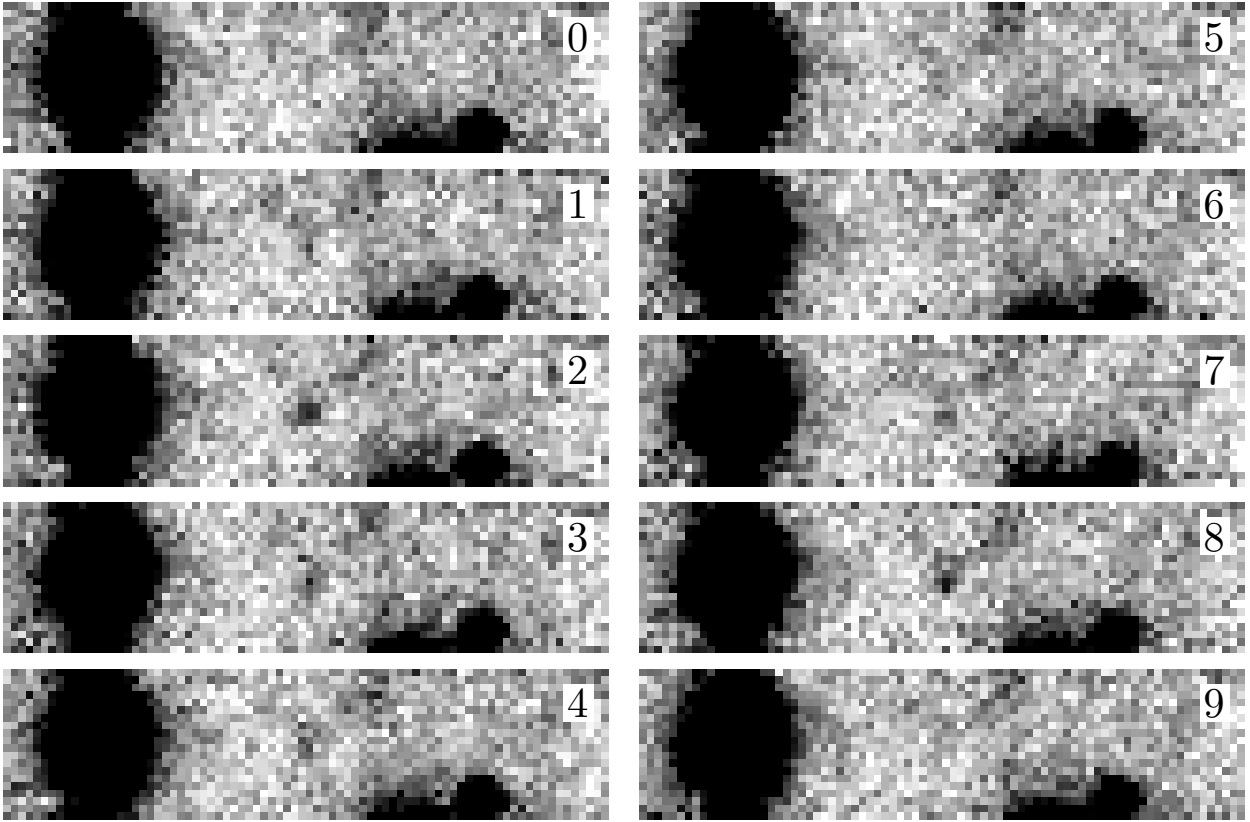}
\caption{Phase-binned total intensity (\ie no polarization information)
images of \psr.
Each image is labeled by its phase bin index (0--9),
with 0 corresponding to phase 0.0--0.1, 1 to phase 0.1--0.2, {\it etc}.
\psr\ is in the exact center of each 40$\times$10 arcsec image.
The intensity peaks in bins 2 and 8.
The measured intensity of the bright star 14 arcsec NE of \psr\
varies by less than 0.2\% over the phase bins.
\label{pulseimg}}
\end{figure}

The integrated image, formed by adding all of the individual images to
eliminate phase and polarization information, is shown in Figure 
\ref{imgtot}.
The total intensity images, formed by combining sets of polarized
images but retaining the 10 phase bins, are shown in 
Figure \ref{pulseimg}.

A 3$\times$3-pixel (1.5$\times$1.5 arcsec) aperture was used
to extract the flux of \psr\ in each phase bin.
While absolute photometric measurements using a small 
(and square) photometric aperture are difficult to calibrate, 
the choice of aperture does not affect the
time-differential photometry, to first order.
The short pulse period (0.385 s) ensures that the point spread function
and image centroids are the same in all phase bins, to first order.
The total intensity of a bright nearby star, measured using the same
analysis routines and photometric aperture used to analyze \psr, is 
measured to be constant to a signal-to-noise ratio exceeding 800.
Assuming the fractional photometric errors are constant, they
contribute no more than a $\sim$ 0.125\% error in the \psr\ measurements.
The individual polarized flux measurements from the nearby bright
star are constant to a similar level ($\sim$ 0.1\%).
Flat-fielding and bias subtraction are performed as described in detail
in \citet{kern2002}.

\subsection{Background Subtracted Pulse Profile\label{background_sub}}

A nearby extended object, which is visible in {\it Hubble Space Telescope} 
WF/PC2 and NICMOS images
of the field around \psr\ \citep{koptsevich_2001}, when observed
from the ground contributes some light to the photometric aperture
used for \psr.
This object is labeled o2 by \citet{koptsevich_2001}.
The \hst\ WF/PC2 images were observed in the F555W filter, a similar bandpass
to that of our ground-based images. 
We convolve the WF/PC2
images with a gaussian PSF whose FWHM equals our average ground-based 
seeing FWHM (1.3 arcsec).
We find that in our 1.5 arcsec square photometric aperture,
the extended object 
contributes 39\% $\pm$ 7\% of the pulse-averaged light in our
(ground-based) photometric aperture.

\begin{figure}
\epsscale{1.0}
\plotone{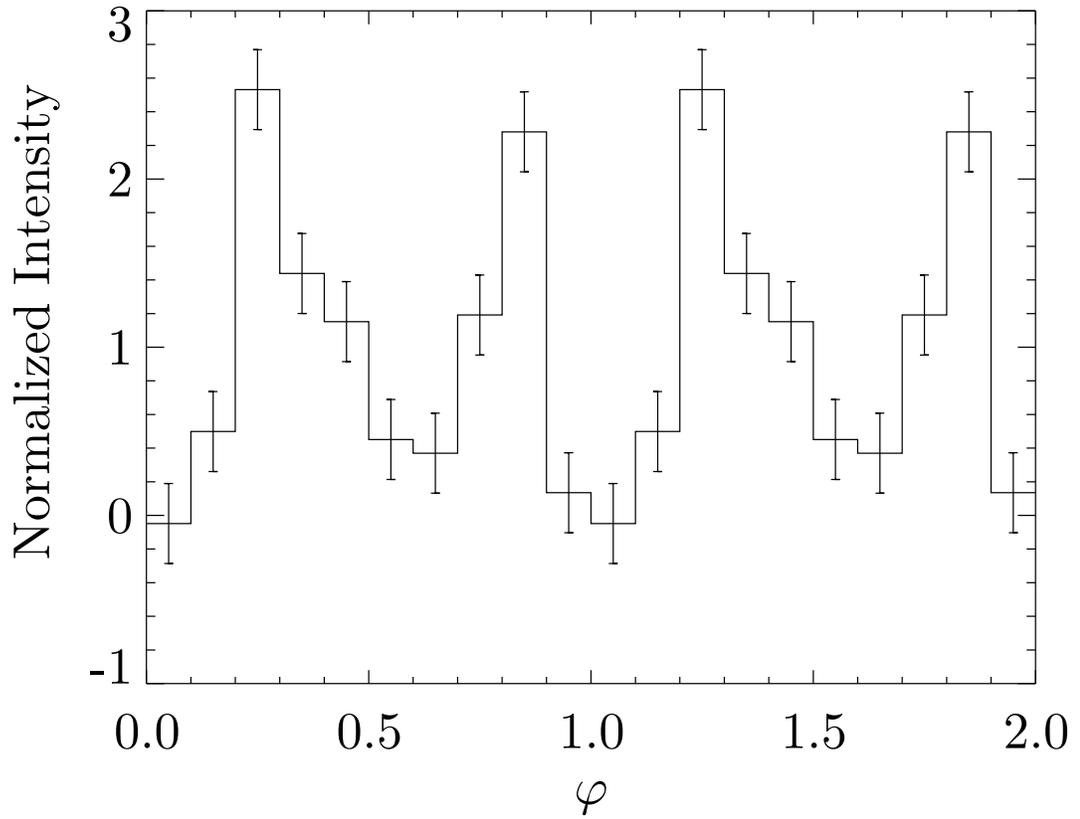}
\caption{Total-intensity pulse profile of \psr.  
The intensity scale is normalized to the pulse-averaged intensity.
Error bars are 1-$\sigma$ errors.
Pulse is plotted twice for clarity.
All intensities have been background-subtracted as described
in Section~\ref{background_sub}.
The radio pulse peaks at phase 0.0, and the possible gamma-ray
peak \citep{ramanamurthy_fkst1996} is at phase 0.2.
\label{pulse}}
\end{figure}

We define $I_i$ as the background-subtracted intensity of \psr\ in
phase bin $i$, which is plotted in Figure \ref{pulse}, normalized
by the pulse-averaged flux, $\overline{I}$.
The uncertainties plotted in Figure \ref{pulse}\ 
are determined from the variations in the
measured intensities in each phase bin, and are only 5\% larger
than the expected Poissonian errors due to the background flux
surrounding \psr.
The phases of our observations are referenced to the peak of the
radio pulse, which arrives at phase 0.0.  
The possible gamma-ray pulse \citep{ramanamurthy_fkst1996} 
peaks at phase 0.2.
The soft X-ray pulse broadly peaks near phase 0.85, 
with a minimum near phase 0.3 \citep{marshall_s2002}.

The minimum of the normalized background-subtracted pulse profile, 
$I_{\rm min}/\overline{I}$, 
is -0.05.
While a negative intensity is clearly unphysical, the measurement
error in each bin ($\sigma_{I} / \overline{I} = 0.24$) 
can account for this discrepancy.
Using all of the bins in a joint probability measurement, we find
a 1-$\sigma$ upper limit (two-tailed) to the true unpulsed flux to be 
$I_{\rm unpulsed}/\overline{I} < 0.16$, accounting for both
measurement error in each bin and the uncertainty in the background
level.
The limited temporal resolution of these measurements causes
$I_{\rm unpulsed}$ to be overestimated, so this number is an upper limit.

The pulse profile 
we measure for \psr\ differs from that
measured by \citet{shearer_1997}.
Due to an ephemeris folding error
(A. Shearer 2002, private communication), the Shearer \etal\
published results must be shifted in phase so that the peak
arrives at phase 0.8 (relative to the radio peak at 0.0).
Shearer \etal\ measure a pulse with a broad single peak, similar
to the soft X-ray pulse profile \citep{marshall_s2002}, reaching
a maximum at phase 0.8, coincident with our Peak 2, but they
find a minimum near phase 0.2, coincident with our Peak 1.
Considering the errors in the flux levels ($\sim$ 80\% of the mean flux) 
and the sky background
level ($\sim$ 40\%) published by Shearer \etal, along with our errors
(24\% flux, 7\% background),
the two curves (independently normalized to the measured mean fluxes)
disagree at the 97.5\% (two-tailed) level.
Some difference may be expected because the Shearer \etal\ results were
obtained through a $B$ filter, while ours were obtained through a wider
passband (400--600 nm).
The measured disagreement could be due to the pulse morphology changing over
the 5 years between the Shearer \etal\ observations and ours,
or simply due to measurements that sample the tail of the error 
distribution.
The level of disagreement does not warrant serious discussion of variability 
in the pulse morphology.

\subsection{Linearly Polarized Flux\label{polfluxsection}}

\begin{figure}
\epsscale{1.0}
\plotone{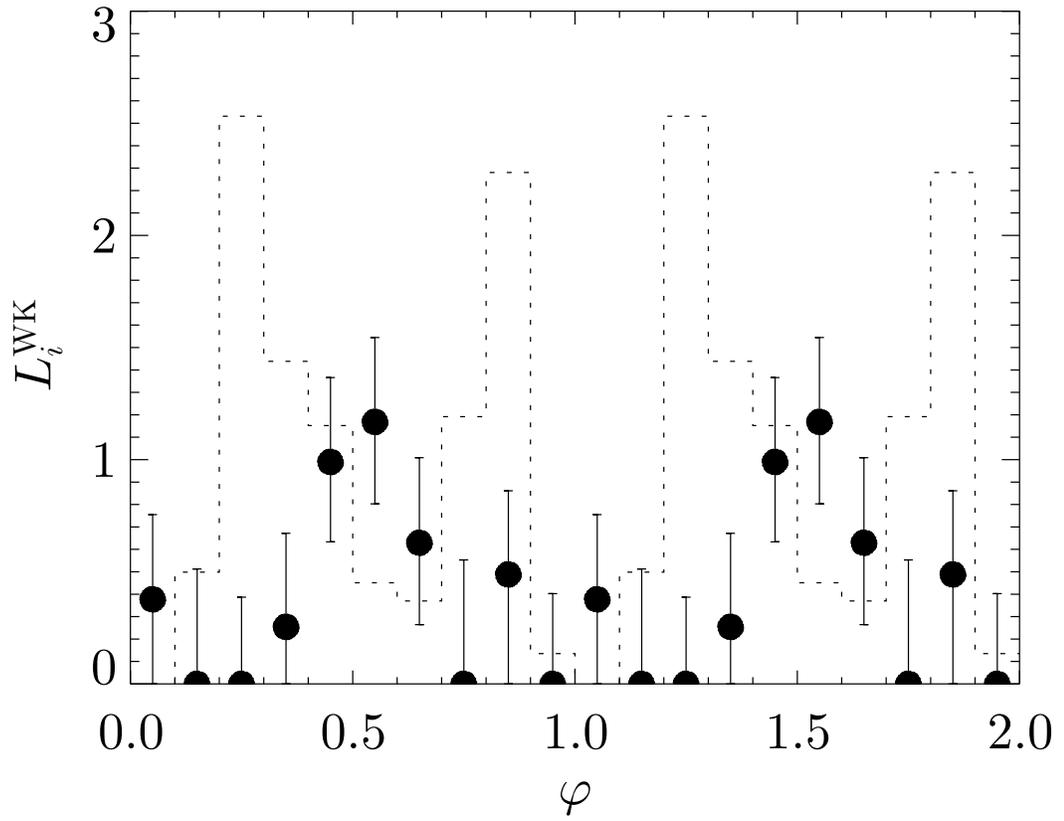}
\caption{Linearly polarized flux.
The filled circles are the Wardle-Kronberg estimates of the
linearly polarized flux, $L^{\rm WK}_i$, 
with 1-$\sigma$ error bars, normalized
to the pulse-averaged total intensity.
The dotted line is the total-intensity pulse profile, on the 
same flux scale, duplicated from
Fig.~\ref{pulse}.
\label{polflux}}
\end{figure}

We compute the linearly polarized flux, $L_i$, and the
polarization position angle, $\theta_i$, from the measured
Stokes parameters $Q_i$ and $U_i$,
\begin{mathletters}
\begin{eqnarray}
L_i & = & \sqrt{Q_i^2 + U_i^2}, \label{eqnpolflux} \\
\theta_i & = & \tan^{-1}(U_i/Q_i) / 2.
\end{eqnarray}
\end{mathletters}
Following the prescriptions in \citet{simmons_s1985}, we construct a
Wardle-Kronberg \citep{wardle_k1974} estimator for $L_i$, which we
denote $L^{\rm WK}_i$, to reduce bias in points with low
signal-to-noise ratios.
A plot of the linearly polarized flux (using the Wardle-Kronberg estimator), 
normalized to the pulse-averaged
flux, is shown in Figure \ref{polflux}.
In phases 0.4 -- 0.7, the $L^{\rm WK}_i$ values differ 
from zero at the 2--3-$\sigma$
level.

We test the significance of the polarized flux measurements in
two ways.
In each test we use the simple estimator in Eqn.~\ref{eqnpolflux}
for $L_i$,
rather than $L^{\rm WK}_i$.
If the uncertainties in $Q_i$ and $U_i$ are distributed as
independent gaussian variables with variance $\sigma^2$, under the null 
hypothesis that there
is no polarized flux, $\chi^2 = \sum_{i=0}^{9}L_i^2/\sigma^2$ 
should be distributed as $\chi^2$ with 20 degrees of freedom.
This first test gives 
$\chi^2 = 50$, which has a cumulative probability of 94.5\%.

The polarized flux measurements form a time series, in which the
ordering of the measurements is significant.
We test the randomness of the time series (\ie, the ordering) 
with the Wald-Wolfowitz 
test of serial correlation \citep{wald_w1943}.
The ordering of the polarization values measured violates the
null hypothesis, that the numbers are randomly chosen, at the 97\% level.

The combination of these two tests, which are independent of one 
another (the size of the measurements is independent of the ordering
of the measurements),
lends some credibility to the measured polarization flux values.
We combine the cumulative probabilities of the
two tests (94.5\% and 97\%) to derive a significance of 0.998
for the detection of pulsed polarized flux (a 3-$\sigma$ result).

If the position angle changes by 90\arcdeg\ on timescales comparable
to the bin width, the measured linear polarization will be reduced.
Therefore, the measured linear polarization must be considered a lower limit.
The best estimate is then that the flux is $\sim$ 100\% polarized
from phase 0.4--0.7, and (linearly) unpolarized at other phases.

\section{Emission Models}

\begin{figure}
\epsscale{1.0}
\plotone{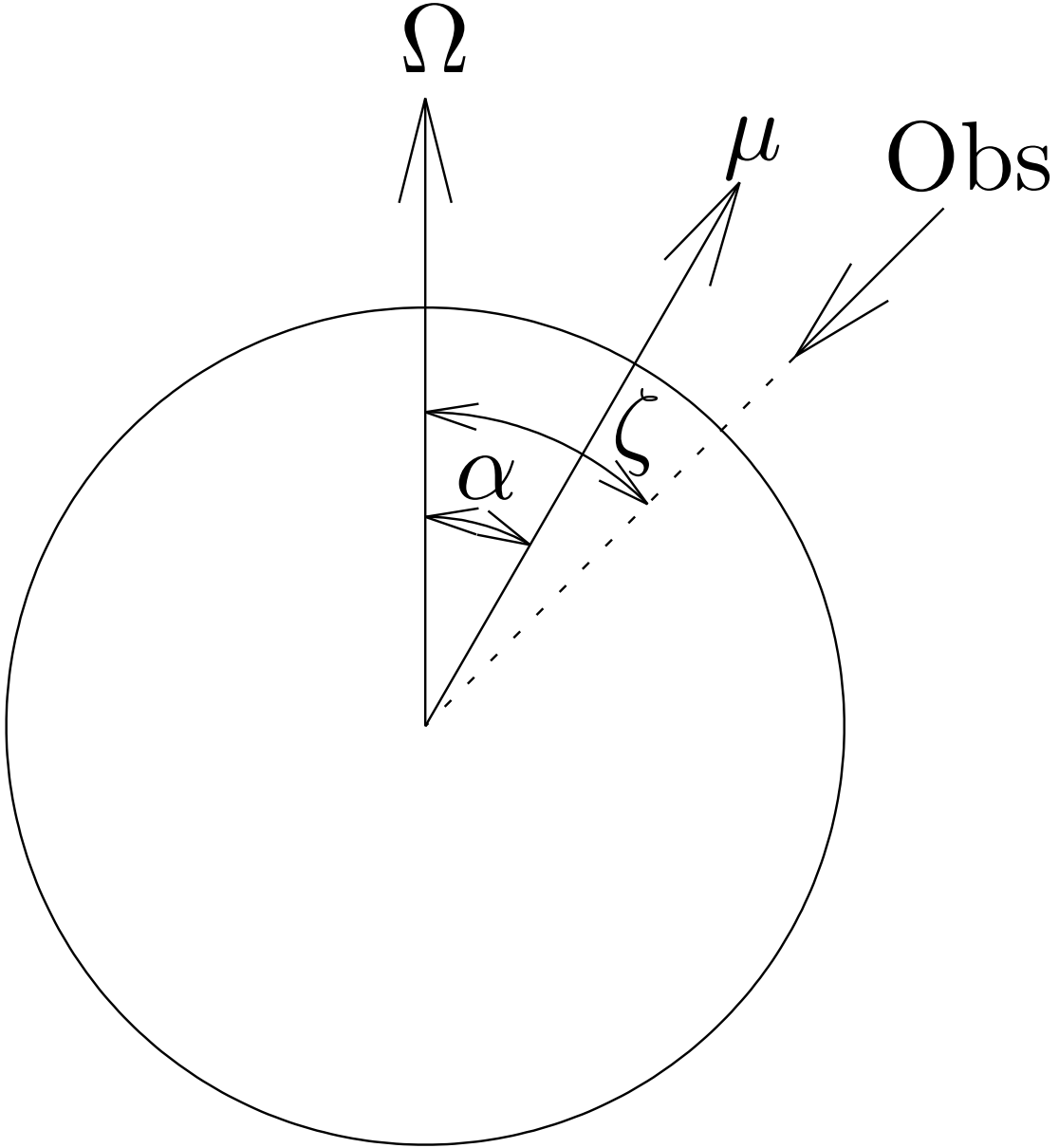}
\caption{Rotation geometry.
$\Omega$ is the rotation axis, $\mu$ is the magnetic
axis, $\alpha$ is the angle between $\Omega$ and $\mu$, and
$\zeta$ is the angle between $\Omega$ and the observer's
line-of-sight.
\label{geom}}
\end{figure}

There are two dominant classes of models that attempt to explain
the origin of the optical emission.
These two classes are the polar cap models and the outer gap models.
In each of these two classes of models, the morphology and the
polarization of the optical light curve are determined by the
spin period, $P$, the angle between the rotation and magnetic axes, $\alpha$,
and the colatitude of the observer's line-of-sight, $\zeta$.
The geometry is shown diagrammatically in Fig.~\ref{geom}.
In addition, the polar cap model allows the height of the emission
region, $h$, to vary.

The optical light curve of \psr\ is sharply double-peaked, unlike
either the radio \citep{gould_l1998} or the X-ray
\citep{marshall_s2002} light curves, which are both 
single-peaked.
We separate the optical light curve (see Fig.~\ref{pulse}) 
into 4 phase intervals.
We define Peak 1 from phase 0.2--0.3, Bridge emission from phase 0.3--0.8, 
Peak 2 from 0.8--0.9, and Off-pulse from 0.9--1.2. 

The radio pulse peaks at phase 0.0,
with FWHM 0.04--0.07 at frequencies from 0.2--1.6 GHz \citep{gould_l1998}.
The Rotating Vector Model (RVM) allows radio polarization profiles to
determine the geometry of the magnetic poles relative to the rotation axis 
and the observer line-of-sight (LOS).
In the case of \psr, the deepest data (at 1.4 GHz) give
a measurement of $\alpha = 29\arcdeg \pm 23\arcdeg$, $\beta = 8.9\arcdeg \pm 6.1\arcdeg$ \citep{everett_w2001}, 
where $\beta$ is the angle of the closest approach of the observer LOS to the
magnetic axis, defined by $\beta = \zeta - \alpha$.
The uncertainties of these angles are highly (positively) correlated.
Two earlier studies investigated \psr, with \citet{lyne_m1988}
giving $\alpha$=8.2\arcdeg, $\beta$=8.2\arcdeg, and \citet{rankin1990}
giving $\alpha$=30\arcdeg\ (with no estimate of $\beta$.)
These earlier studies do not estimate errors, as they are fits
to empirical assumptions about the underlying geometry whose errors
are not easily estimated.

\subsection{Polar cap model}

The polar cap model assumes that the optical light is emitted
near the surface of the neutron star, at
radii small compared to the light cylinder radius.
The optical light curve and polarization information can be
tested against two predictions of the polar cap model.
First, the polarization position angles can be compared to the
predictions of the RVM, using geometric parameters determined
by radio polarization data.
Second, the phase separation between the peaks in the optical light
curve, combined with the geometric parameters determined by
the RVM, determines an emission height for the optical peaks.
If this emission height is too large to be consistent with
the polar cap model, then the model fails.

The RVM is a simple model, which assumes that the polarization
position angle depends only on the projection of the magnetic
dipole axis on the sky.
Because the position angles predicted by the RVM have no dependence on 
height above the neutron star surface,
we can reasonably assume that the radio and optical position angles
must follow the same fit, even if they are emitted in different regions.
The radio polarization data of \citet{everett_w2001} are not
calibrated with an absolute position angle, which leaves the
zero-point of radio position angles as a free parameter.
We take the RVM position angles predicted by the fits to the radio
polarization, degrade the temporal resolution to that of the
optical observations (10 resolution elements),
and add a constant position angle offset to best fit the optical polarization
measurements.
The best-fit RVM prediction, showing both the radio and optical
position angles, is shown in Figure \ref{rvm}.
The range of $\alpha$ and $\zeta$ allowed by the radio fit
does not change the predicted position angle
by an amount comparable to the errors in our measurement
of $\theta_i$, so the optical data do not reduce
formal errors on $\alpha$ and $\zeta$ beyond those of the 
radio polarization
data alone.
The fit of the optical position angles to the RVM prediction
is quite good, with a
measured $\chi^2$ of 2.3 for 2 degrees of freedom
(cumulative probability 0.69).
The optical position angle observations
greatly expand the phase coverage of the polarization
data, but the size of the optical error bars
and the lack of absolute calibration of the
radio data angles prevent significant improvement of
the RVM fit.
The fact that the optical data agree well with the
radio data, given that they are obtained by entirely
independent techniques, does lend credibility to the
RVM.
The conservative conclusion drawn from the the optical polarization 
position angles 
is that they are consistent with the assumption that the optical light is
emitted along dipolar field lines, without consideration
of relativistic aberration, light travel time, or retarded potentials.

The polar cap model, by assuming that the light is emitted
near the neutron star surface (far from the light cylinder), 
makes specific predictions
regarding the morphology of the light curves.
For the second test, the information contained in the optical
pulse profile is reduced to the separation between the peaks,
which are assumed to be emitted from a single height, $h_{\rm peak}$,
above the neutron star surface.
The last open field lines form a pseudo-cone near the surface of the 
star, and the optical light is assumed to be emitted along the
surface of the cone (tangent to the field lines), 
from some height $h_{\rm peak}$ above the surface.
Peaks in the optical light curve arise as the neutron star rotates 
and the line-of-sight intersects the cone of emission at 
$h_{\rm peak}$.
The sharp rise of Peak 1 and sharp fall of Peak 2 match the
prediction that the optical emission arises from a hollow cone
\citep{sturner_d1994},
with the radio pulse arriving in the center of the cone,
during the optical Off-pulse phase interval.

Several theoretical attempts have been made to estimate the height of
the gamma-ray emission region.
Recent studies have estimated that particle acceleration
begins at heights 0.5--1 times the neutron star radius,
$R_{\rm NS}$, above the poles
\citep{harding_m1998b}.
This particle acceleration results in radiation of curvature
photons or inverse Compton scattering of thermal photons,
which then produce pairs, resulting in cascades of photons 
and particles.
The optical radiation is produced after some number of
generations of this cascade.
We are not aware of specific estimates of the height of the
optical emission regions with respect to the primary particle
acceleration regions, but it has been speculated that
the gamma-ray emission regions may extend to several
($\sim$ 3--5) $R_{\rm NS}$ \citep{daugherty_h1996a}, 
motivated by the requirement that observed
gamma-ray pulsar statistics reflect random viewing angles of
pulsars.
We adopt the estimate that $h_{\rm peak} \lesssim 5 R_{\rm NS}$.

The opening angle of the cone defined by the last open field lines
increases with increasing height, $h_{\rm peak}$, of the emission region
above the neutron star surface.
All else equal, larger values of $h_{\rm peak}$ result in more widely 
separated peaks in the optical light curve.
Given values of the pulsar period, $P$, and the phase
separation between the optical peaks, $\Delta\varphi_{\rm peak}$,
$h_{\rm peak}$ is a function
of $\alpha$ and $\zeta$.
We calculate the relationship between these variables by 
assuming the magnetic fields are described by an inclined
static dipole (defined by $\alpha$), 
and define the last open field lines as those
which are tangent to the light cylinder (defined by $P$).
We do not consider plasma loading, inertial frame dragging,
or general relativistic photon propagation effects near the surface, 
and do not distort the field lines for any near-surface plasma
effects.
These additional corrections, which we neglect, can increase
the opening angle of the near-surface field lines by a factor
of $\sim$ 2 \citep{daugherty_h1996a}.
The absence of these corrections results in conservative limits,
in the sense that the corrections would allow larger values of
$\alpha$ and $\zeta$ to produce the observed peak separation.
For given values of $h_{\rm peak}$ and $\zeta$, 
we calculate the tangents to the last
open field lines, and find the values of $\varphi$ for which
the tangents to the last open field lines point toward the observer.
Because $\alpha$ and $\zeta$ are not uniquely determined
for \psr, estimates of $h_{\rm peak}$ depend on the range of $\alpha$
and $\zeta$ allowed by the RVM.
We do not use the standard small-angle approximations for the
dipole field lines, but calculate the full spherical trigonometric
relationships for all angles, which are valid for all values of
$\alpha$ and $\zeta$.

By assuming that the optical emission forms a hollow cone
\citep{sturner_d1994},
with the radio pulse in the center of the cone, the separation
between the optical peaks becomes 
$\Delta\varphi_{\rm peak} = 0.4$ (instead of 0.6).
A plot of $h_{\rm peak}$ versus $\alpha$ and $\zeta$, showing values
of $h_{\rm peak}$ comparable to 
several $R_{\rm NS}$, is shown in Fig.~\ref{heightsmall}. 
It must be recognized that the uncertainties in $h_{\rm peak}$ for
each $\alpha$ and $\zeta$
amount to several tens of percent, in the sense that the
plot is calculated for $\Delta\varphi_{\rm peak} = 0.4$, which is itself
uncertain by approximately $\pm0.1$.
If $\alpha$ and $\zeta$ are less than 
$\sim 5^{\circ}$, $h_{\rm peak}$ is small 
enough to be consistent with the intial 
assumption of the polar cap model (that the emission region
is within a few NS radii) and the
constraints of the RVM.

The data contained in Fig.~\ref{heightsmall}\ represent only
a consistency check based on the optical peak separation,
from a single height above the neutron star surface.
We propose an entirely empirical model to estimate the 
relative emissivity versus height in the polar cap model,
for a range of heights.
For a single representative point in Fig.~\ref{heightsmall},
we calculate the emission height at all phases, defining the height as the
point at which the tangent to the last open field lines points to
the observer.
A plot of the emission height versus $\varphi$ for $\alpha = 3.5^\circ$, 
$\zeta = 5^\circ$ is shown in Fig.~\ref{hvsphase}.
While this choice of $\alpha$ differs somewhat from the best-fit RVM estimate
of $29\pm23^\circ$ from radio data, it corresponds to
$h_{\rm peak} ~ 5 R_{\rm NS}$, which we choose as an upper limit.
For this analysis (but not in the following Outer gap model section),
we make use of the lag between the radio peak and the 
transit of the magnetic pole, as determined from the radio
polarization data in \citet{everett_w2001}.
They find that the RVM estimates a lag of $14.9\pm0.9^\circ$ of the
magnetic pole with respect to the radio peak, 
which we approximate as $\Delta\varphi$ =  0.05 (or 18$^\circ$).
We find this lag convenient, as it makes the interpretation of
the pulse profile (Fig.~\ref{pulse}) symmetric, in the sense
that the first peak arrives at $\varphi \sim 0.25$ 
(between 0.2 and 0.3 relative to the radio peak),
which is 0.2 revolutions after the magnetic
pole transit, and the second peak arrives at $\varphi \sim 0.85$, 
which is 0.2 revolutions before the magnetic pole transit.
This should not be interpreted as an independent corroboration 
of the estimated lag of the magnetic pole, as there
is no reason to believe the profile must be symmetric,
and because the effect is smaller than the optical temporal resolution.

The emissivity at different heights above the polar caps
can be crudely estimated using the height profile shown
in Fig.~\ref{hvsphase}.
We again assume light is emitted along tangents to the last open
field lines.
Points on the two-dimensional surface of last open field lines emit light
that would be observed at different phases, $\varphi$, by observers
at different colatitudes, $\zeta$.
The transformation from two-dimensional area on the surface of
last open field lines to solid angle ($\varphi$, $\zeta$) on the sky into which
optical light is emitted defines
a Jacobian, which determines the relationship between 
emissivity (emitted flux per unit surface area) and 
the flux observed in a given phase interval (recognizing that
the observer defines a delta-function in $\zeta$).
At lower heights, the magnetic field lines diverge
more rapidly, which results in a lower observed flux for a given
emissivity (\ie, the flux is emitted into a larger solid angle).
Conversely, given an observed flux, the inferred emissivity is
greater at lower heights.

A plot of the estimated emissivity versus height above the
neutron star surface is shown in Fig.~\ref{emissivity},
for $\alpha = 3.5^\circ$, $\zeta = 5^\circ$.
The points at small $h$ (less than $\sim 3 R_{\rm NS}$) come from
the Off-pulse interval, the peaks
in emissivity correspond to the optical peaks, 
and the largest value of $h$ corresponds to 
$\varphi$ = 0.5--0.6.
This plot clearly shows the errors implicit in Fig.~\ref{heightsmall},
based on the size of the horizontal bars (the range of heights that
contribute to each phase bin).
The heights and emissivities shown in Figs.~\ref{hvsphase} and 
\ref{emissivity}\ depend on the
given geometry ($\alpha = 3.5^\circ$, $\zeta = 5^\circ$), 
but the trends
in the plots remain the same for different geometries, with a 
rapid rise of emissivity up to $h_{\rm peak}$, declining
at larger heights.
It must be emphasized that this emissivity model is purely empirical, 
and is not based
on any understanding of the underlying pair cascade physics
or emission mechanisms.

\subsection{Outer gap model}

Outer gaps arise beyond null charge surfaces, where the
Goldreich-Julian charge density vanishes \citep{romani_y1995, romani1996}.
These gaps act as accelerators, which result in cascades of
high-energy particles and photons being emitted along the
magnetic field lines (in the rotating frame).
The gaps are ``closed'' some distance from the null charge surface,
where the cascades supply a sufficient 
density of charged particles to end the acceleration.
In a broad interpretation, the regions responsible for the observed 
optical emission
are bounded by the null charge surfaces, the last open field lines,
the light cylinder, and the upper edges of the gap (where the
cascades close the gap).
Individual models, applied to different pulsars, have assumed
that high-energy emission comes from some subset of the 
potential emission region.

The typical assumption used to model the observed emission
from an outer gap is that photons are emitted, in the rotating
frame, along magnetic field lines.  
Calculating the resulting light curve entails
applying relativistic corrections and correcting for light
travel times, which means that the emission region observed
at a particular phase is generally not along the line-of-sight
connecting the observer to the surface of the neutron star.
In addition, calculating the magnetic field line trajectories
at radii comparable
to the light cylinder radius requires using retarded
potentials, such as those described by \citet{deutsch1955}.

The parameters required to produce an individual light curve using
an outer gap model are $P$, $\alpha$, and $\zeta$, as in the polar
cap model, plus a width parameter $w$.
The width parameter, $w$, identifies a set of magnetic field
lines, based on the location at which they intersect the
surface of the neutron star.
The last open field lines are defined to lie at $w=0$, while
the magnetic pole defines $w=1$ \citep{romani_y1995}.
A given value of $w$ identifies a two-dimensional surface
(a pseudo-cone), that is the collection of all magnetic field lines
that strike the neutron star surface at the appropriate distance from
the magnetic pole.
We then assume that this surface emits
optical photons tangential to the local field lines,
directed inward, outward, or both, 
with a constant flux per unit surface area.
More complex outer gap models may also include azimuthal
and radial limits on the emission region inside the outer gap,
or any arbitrary pattern of emission over the surfaces of
constant $w$.

\begin{figure}
\epsscale{1.0}
\plottwo{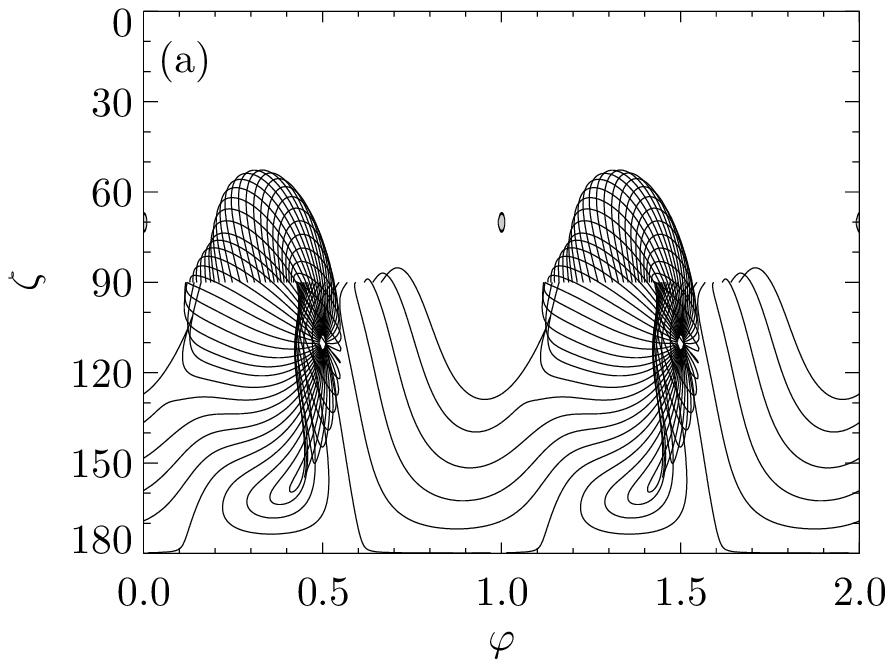}{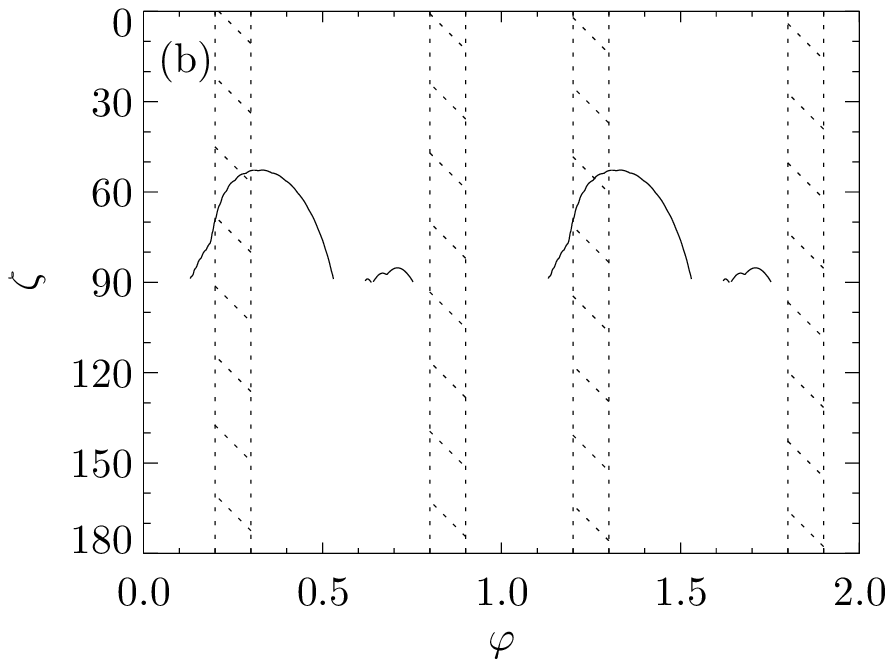}
\caption{(a) Outward-going emission skymap for $\alpha=70^\circ$, $w = 0.20$.
Each point in the plot represents the phase, $\varphi$, at which
an observer at colatitude $\zeta$ would observe an outward-going
photon emitted along magnetic field lines in the $w=0.20$ 
surface.
Only values of $\zeta < 90^\circ$ are in the outer gap (the
null-charge surface corresponds to $\zeta=90^\circ$), where photons
are produced.
The small circles at $\zeta=70^\circ$, 
$\varphi=0$, 1, and 2 represent the magnetic pole 
responsible for the radio emission.
(b) Envelope of outer gap emission region.
The lines denote the boundary between areas with observable emission
and those without.
These lines include virtually all peaks in the light curves.
The hatched regions show the locations of the optical peaks
(relative to the radio peak at $\varphi=0$).  
The envelope does not intersect the region $\varphi$ = 0.8--0.9 for
these parameters, implying that this geometry cannot produce
the observed optical peaks.
\label{skymap}}
\end{figure}

In our simplified approach, where we only determine emission
peaks for a given geometry, we adopt a simple technique for 
identifying peaks.
For a given $\alpha$ and $w$, a skymap is constructed,
as described in \citet{yadigaroglu1997}.
A skymap shows the locus of points in the $\zeta$-$\varphi$ plane
(where $\varphi$ is phase) which correspond to observable emission
from the surface of magnetic field lines defined by $w$.
The skymap corresponding to $\alpha = 70^\circ$, $w = 0.2$ is
shown in Fig.~\ref{skymap}.
The transformation that maps the two-dimensional surface of magnetic
field lines at constant $w$ to the two-dimensional skymap ($\zeta$
vs.\ $\varphi$), combined with the assumption of constant emitted flux
per unit surface area, results in a map of observed flux per
unit $\zeta$-$\varphi$ area. 
The Jacobian of this transformation is large (high flux) where
$\zeta$ and $\varphi$ change little across or along magnetic field
lines.
This is most commonly the case at the envelope of the optical
emission regions, between regions of $\zeta$ and $\varphi$ where 
there is observable emission and regions where there is none.
We then use the location of this boundary as an estimate of the
$\zeta$ and $\varphi$ at which the optical light curve peaks.
One motivation for this formulation is that for a given
$\zeta$, the lightcurve will be discontinuous at values of $\varphi$
which cross this boundary, as the flux is zero on one side of
the boundary.
While this quality does replicate part of the phenomenology 
associated with a peak, by only reproducing the positions
of the optical peaks, we make no distinction between zero 
emission off-peak and bridge emission.

This simplified technique is prone to identify too many peaks.
There are configurations for which the outer gap models generate
broad, sinusoidal pulse profiles, whose peaks are not captured by 
identifying boundary crossings.
In addition, there are low flux regions identified by
this technique that are insignificant in the overall
pulse profile.
Configurations with small values of $\alpha$, or large
values of $\zeta$ (near 90$^\circ$) tend to
produce more spurious peaks using this technique.

Given the large number of pulse profiles that can be constructed
by varying the available parameters, we restrict the
observational features we try to match.
We reduce the contents of the 
optical light curve to the peak positions,
specifically, Peak 1 occurs between phases 0.2--0.3 and 
Peak 2 occurs between phases 0.8--0.9.
The radio pulse arrives at phase 0.0.
We model the emission from three classes of outer gap models, to match the 
observations.
Class A restricts the emission region to single values of $w$, and 
insists that the phases of the two optical peaks and the radio
peak all match the observations.
Class B allows emission from all values of $w$, and insists
that the phases of the radio and optical peaks match.
Class C allows emission from all values of $w$, but assumes
the optical peaks have no relationship to the radio peak.
For each of these classes, we determine the light curves
produced by only outward-going photons, only inward-going
photons, and a combination of inward- and outward-going photons.
For each of the nine resulting constraints, we determine
the values of $\alpha$, $\zeta$, and (for Class A) $w$ which 
produce the observed phases of optical peaks.

We calculate skymaps, and their corresponding envelopes, for
$\alpha$ = 5$^\circ$--85$^\circ$ at intervals of 5$^\circ$, and
for $w$ = 0.02, 0.05, 0.10, 0.20, 0.40, and 0.80.
From this grid of skymaps, we calculate peak positions for
values of $\zeta$ = 1$^\circ$--90$^\circ$ at intervals of 1$^\circ$.
The agreement of the models with the constraints corresponding to
each class, and inward- or outward-going emission, are
recorded and binned into 5$^\circ$-square bins in $\alpha$-$\zeta$
space.
The results of the agreement of the models is presented in
Fig.~\ref{outgap}.
We include the regions of $\alpha$ and $\zeta$ which are
consistent with the RVM applied to the radio data in Fig.~\ref{outgap},
except in Class C, which ignores the radio data. 

Further work will be required to compare the observed polarization
profile to those predicted by the outer gap models under
different geometries, in an attempt to further constrain the
outer gap models.

\section{Discussion}

The double-peaked nature of the optical pulse profile can
be interpreted in light of both polar cap and outer gap models.
In the absence of a clear gamma-ray pulse profile for \psr,
and given that the soft X-rays observed for \psr\ are
commonly interpreted as thermal emission from the neutron star
surface,
the optical pulse profile is the only high-energy data available
to test these models.
The optical polarization data, while not of high signal-to-noise ratio,
allow a new set of diagnostics to be applied to these tests.

Unlike the radio emission, which is probably coherent curvature
radiation, the optical emission mechanism is more likely
to be synchrotron radiation.
This is supported in the case of \psr\ by the close match between the optical
fluxes and the extrapolation of the non-thermal
X-ray power law component,
scaling as $\nu^{-0.45}$ \citep{koptsevich_2001}.
There are two immediate implications of this
inference.
First, the polarization position angle differs by
90$^\circ$ between curvature and synchrotron radiation.
Curvature radiation in the normal mode is
polarized parallel to the $B$ field, while
the position angle of syncrotron radiation will be
perpendicular to the $B$ field.
We cannot test this prediction without
absolutely calibrated radio position angles.
Without absolutely calibrated position angles, 
there is no orientation against which to compare
the measured optical position angles.
Second, if the optical light is synchrotron with a negative
power-law exponent, the
$B$ field at the location of emission must be low enough
that the synchrotron characteristic frequency is below
the optical frequency.
A critical frequency, $\nu_{\rm c}$, of $ \lesssim 10^{14}$ Hz corresponds to 
$B \lesssim 10^{8}$ G, compared to the surface $B_0 = 5 \times 10^{12}$ G.
This constraint leads to a minimum height of the emission
region of $\gtrsim 35 R_{\rm NS}$ for the dipole field to drop to 10$^{8}$ G.

\begin{figure}
\epsscale{1.0}
\plotone{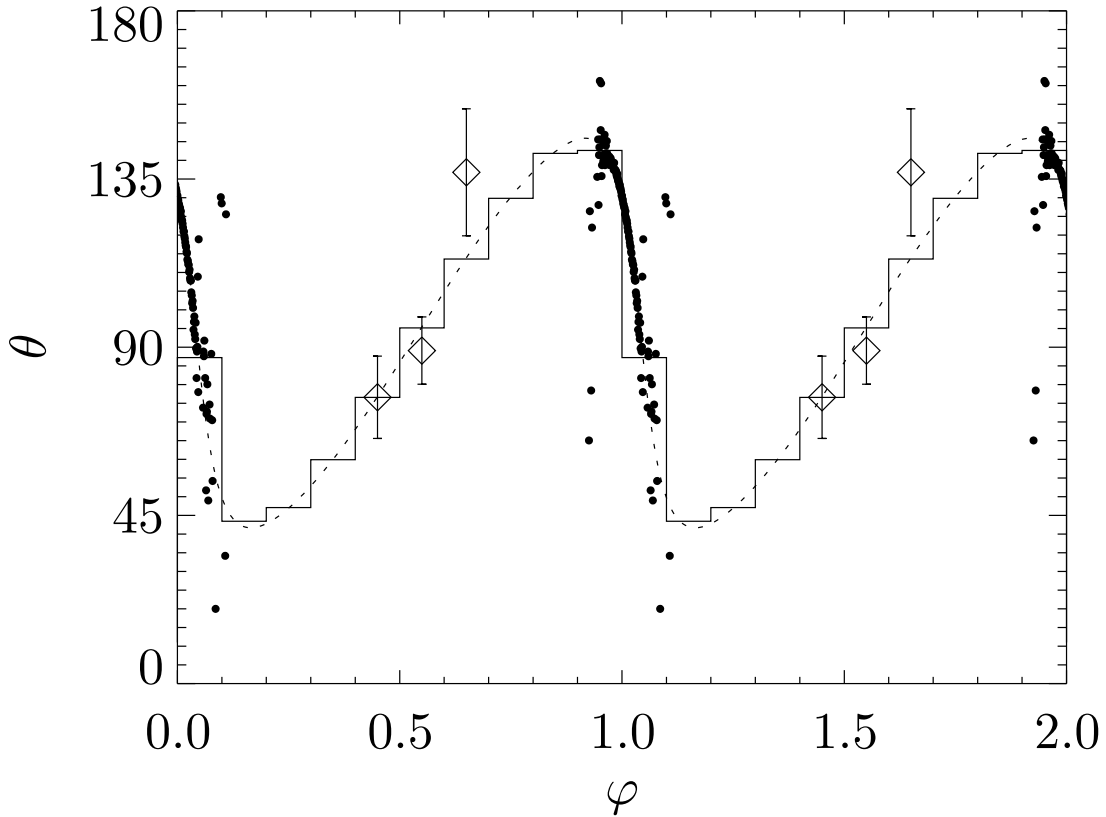}
\caption{Linear polarization position angles, $\theta$.
Diamonds are the optical position angles with 1-$\sigma$ error bars,
dots are the radio polarization position angles from \citet{everett_w2001}.
The radio position angles are not absolutely calibrated, and are shown
with the best-fit zero-point offset.
The radio position angles do not have error bars plotted.
The radio position angle measurements made where the radio flux
is low ($|\varphi| > 0.05$) have large uncertainties.
The dotted line is the prediction from the Rotating Vector Model,
using the radio data.
The solid line is the same prediction, with temporal resolution 
reduced to equal that of
the optical measurements.
\label{rvm}}
\end{figure}

\begin{figure}
\epsscale{1.0}
\plotone{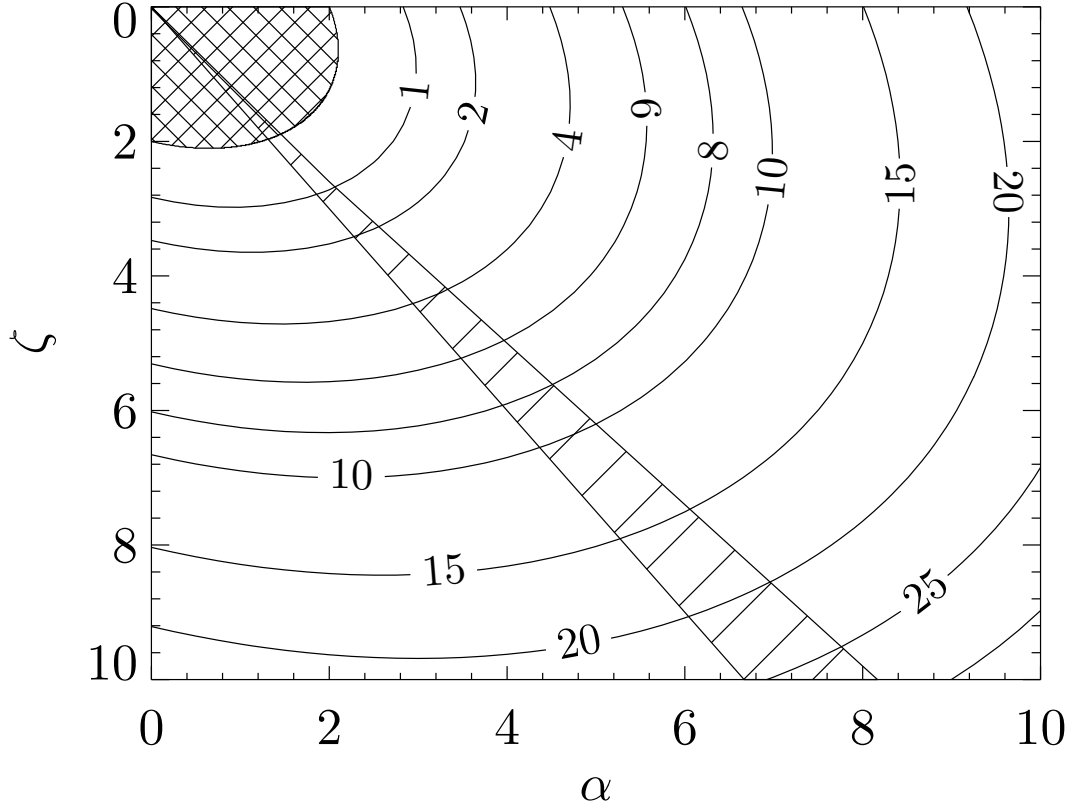}
\caption{Height at which optical peaks are emitted.
Contours represent heights of the optical emission region,
above the surface of the neutron star, that give the
observed phase separation between peaks of 0.4.
The heights are stated in multiples of 
$R_{\rm NS} = 10$ km.
The hatched region is the 3-$\sigma$ confidence region for
$\alpha$ and $\zeta$ (no dependence on $h_{\rm peak}$), 
applying the Rotating Vector Model
to the radio data of \citet{everett_w2001}.
The cross-hatched region denotes $h_{\rm peak} < 0$, for which there is
no geometry giving the observed peak separation.
This plot shows only values of $\alpha$ and $\zeta$ 
corresponding to a range of $h_{\rm peak}$ comparable to
the theoretically expected heights.
\label{heightsmall}}
\end{figure}

\begin{figure}
\epsscale{1.0}
\plotone{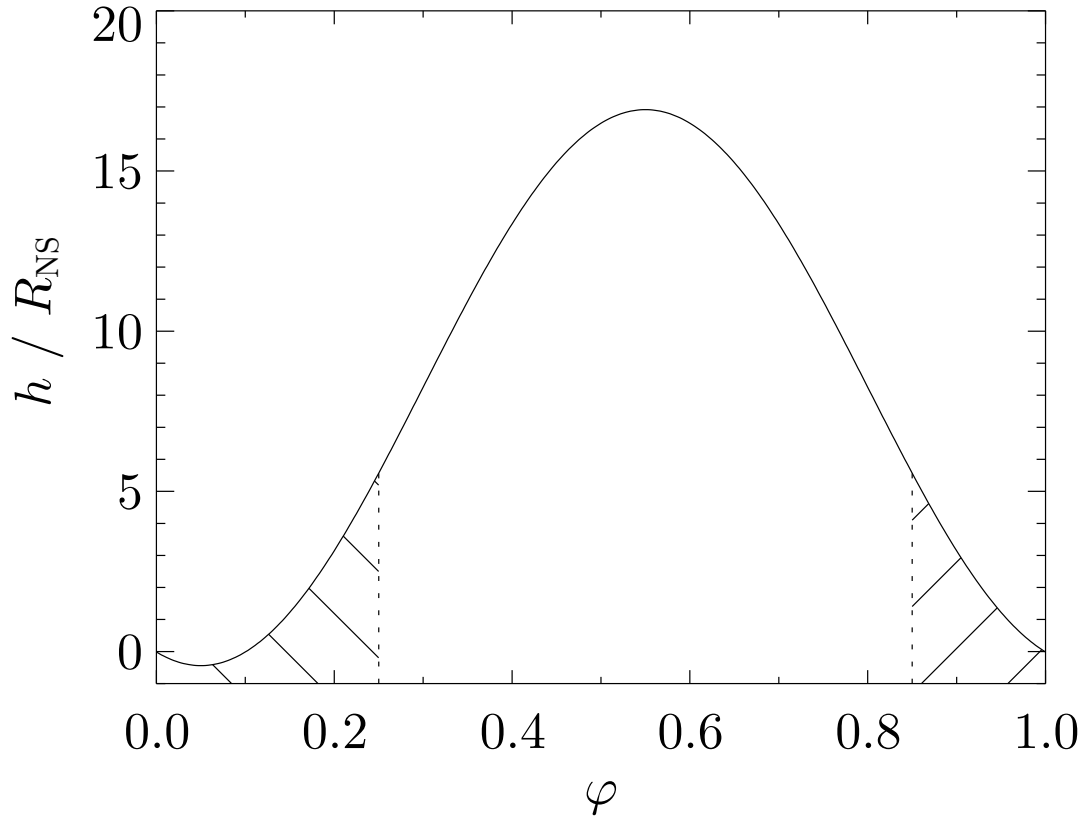}
\caption{Sample optical emission height profile in polar cap model.
This plot is calculated for $\alpha = 3.5^\circ$, $\zeta = 5^\circ$
(see Fig.~\ref{heightsmall}),
assuming the magnetic pole transit lags the radio peak by 0.05 in phase.
The height, $h$, at which the tangent to the last open field lines point
to the observer, is plotted versus phase, $\varphi$, normalized to
the neutron star radius, $R_{\rm NS}$ = 10 km.
The vertical dotted lines correspond to the nominal locations of
the optical peaks (during the phase intervals 0.2--0.3, and 0.8--0.9),
and the hatched regions approximately mark the Off-pulse interval.
\label{hvsphase}}
\end{figure}

\begin{figure}
\epsscale{1.0}
\plotone{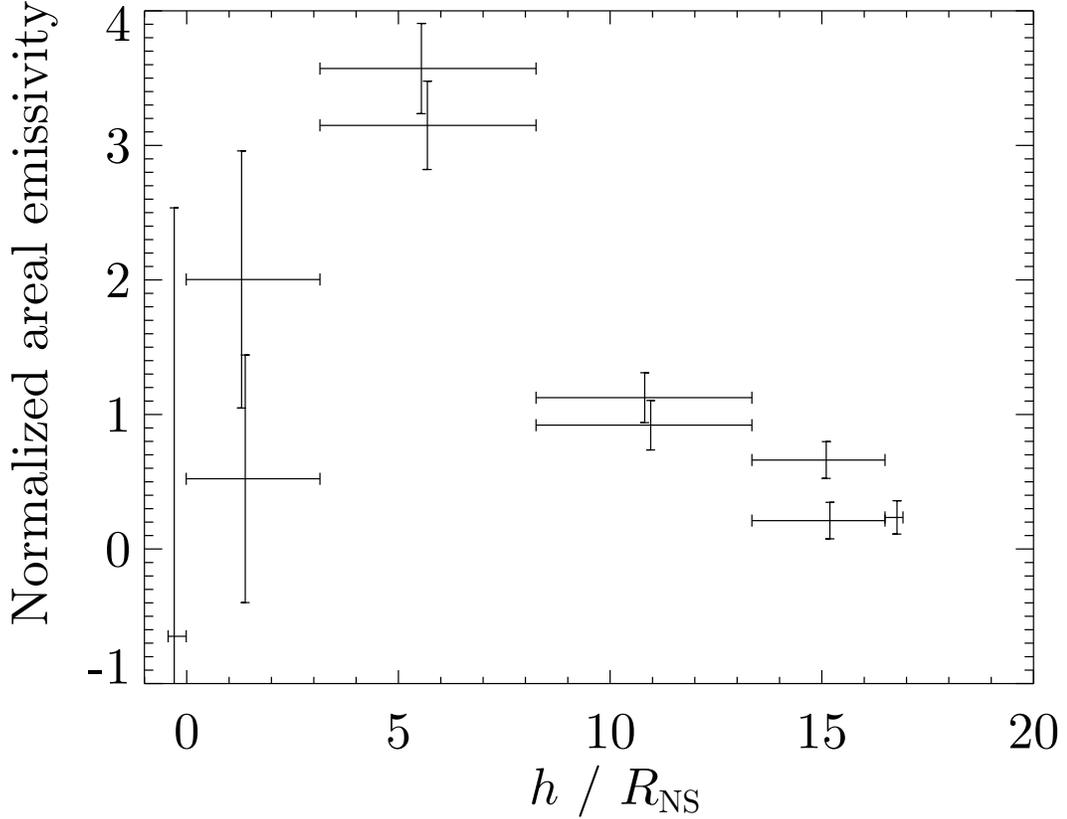}
\caption{Sample areal emissivity versus height in polar cap model.
The emissivity is calculated from the observed pulse profile 
(Fig.~\ref{pulse}) and
the Jacobian of the transformation from area on the last closed
field lines (at the emission location as defined in Fig.~\ref{hvsphase})
to solid angle on the sky.
The geometric assumptions are as in Fig.~\ref{hvsphase}, and
the plotted values vary as $\alpha$ and $\zeta$ vary.
Ten points are plotted, corresponding to the ten measured points
in the pulse profile.
The vertical error bars are the 1-$\sigma$ errors from Fig.~\ref{pulse},
the horizontal bars denote the range of heights associated with
the phase interval in each phase bin.
\label{emissivity}}
\end{figure}

The polar cap model can explain the morphology of the optical pulse 
profile, with the light arising from an emission within
$\sim 5 R_{\rm NS}$, for $\alpha$, $\zeta \lesssim 5^\circ$,
as shown in Fig.~\ref{heightsmall}.
As is typical for polar cap models with widely separated
peaks \citep{sturner_d1994}, 
the predicted values of
$\alpha$ are small.
What is striking is that the polarization position angles
for the optical light curve agree very well with the sweep
predicted by the Rotating Vector Model, applied to the radio
polarizaion data of \citet{everett_w2001}.
While the agreement is quite good ($\chi^2$ = 2.3 for 2
degrees of freedom), it is based on only three data points,
which give only two degrees of freedom because the absolute position angles of
the radio data are not known.
The optical position angles do support the interpretation of
the radio data in \citet{everett_w2001}, which is encouraging,
as the fits to the RVM using radio data from a short phase interval
give estimates with a great deal of
covariance in the parameters.
The simple fact that the sweep of optical position angles agrees
at all with the radio position angles implies a new level of
confidence that the RVM applies to \psr, considering that the RVM
gives good fits to only a small fraction of pulsars with
high-quality radio polarization data.
It must be recognized that the radio polarization data are obtained 
in only a small window of phase (see Fig.~\ref{rvm}), 
while the optical position angles are taken from a wider range of
phase.
The optical data, when combined with the radio data, give a
more complete picture than is available for nearly any other
radio pulsar, except the Crab (which does not follow the RVM).
While the optical data do not directly confirm the 
estimated lag between the radio peak and the magnetic pole
transit, the agreement of the optical data with the RVM fit
does bolster confidence that the RVM can be interpreted 
quite literally in the case of \psr.
The optical position angles do not allow us to test if
the optical light is synchrotron while the radio
is curvature radiation, because the radio position angles are not
absolutely calibrated.
Regardless of the interpretation of the optical position
angles, the polar cap model can explain the general morphology
of the optical light curve (peak positions relative to the
radio peak, bridge emission between the peaks).
However, there may
be a problem with an unbroken synchrotron
power-law spectrum through optical frequencies requiring a larger
emission height ($\gtrsim 35 R_{\rm NS}$) 
than is realistic under polar cap cascade scenarios ($\lesssim 5 R_{\rm NS}$).

\begin{figure}
\epsscale{0.8}
\plotone{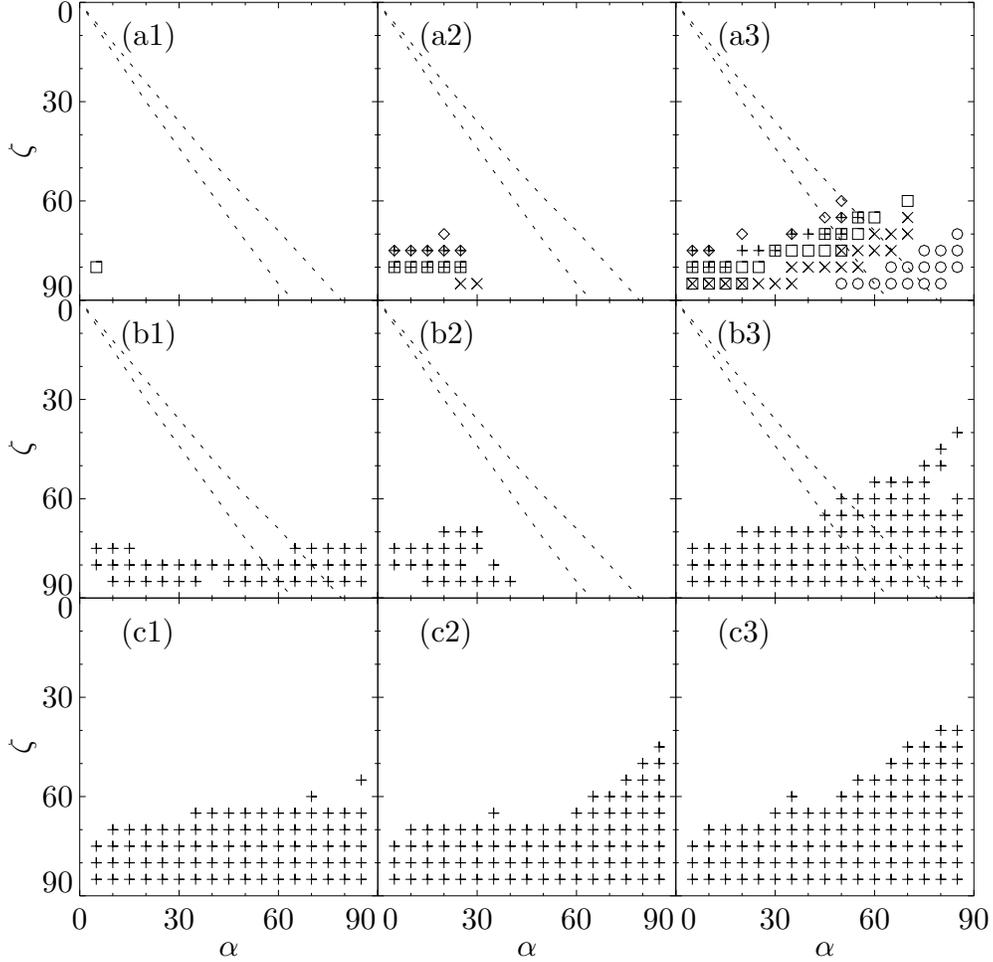}
\caption{Outer gap parameters which match observed optical properties.
Each panel is marked at values of $\alpha$ and $\zeta$ for which 
the outer gap models can produce a light curve with peaks at the
observed phases.
The (a) panels produce peaks at phases 0.2 and 0.8 relative to 
the radio peak at phase 0.0, from field lines at a single value of $w$.
The (b) panels produce peaks at phases 0.2 and 0.8 relative to 
the radio peak at 0.0, from field lines at all values of $w$.
The (c) panels produce peaks separated by 0.4 (or, equivaliently, 0.6) 
in phase, with no 
specified relationship to the radio peak, and from all $w$.
The (1) panels show only outward-going emission, 
the (2) panels only inward-going emission, 
and the (3) panels both inward- and outward-going emission.
In the (a) panels, the symbols represent the value of $w$ 
satisfying the observations, with a diamond for 0.02, 
plus for 0.05, square for 0.10, cross for 0.20, and a circle for 0.40.
No models with $w=0.80$ fall on this plot.
The sole data point in panel (a1) is likely to be spurious.
The region in the (a) and (b) panels 
marked by the dotted lines is the 3-$\sigma$ confidence
interval determined from the RVM fits to radio data.
\label{outgap}}
\end{figure}

The outer gap models discussed here are somewhat restrictive,
in that we do not accomodate several dimensions of flexibility
available to outer gap models.
We assume a constant emitted flux per unit surface area across
the entire possible outer gap region.
This assumption ignores, for example, the flexibility allowed by instituting
azimuthal or radial limits to the optical emission region,
which could shift the location of the peaks in the optical light
curve.
We also reduce the optical light curve to its simplest observable
parameter, the location of the optical peaks.
We find that with these restrictions, we are able to find
agreement between the data and the outer gap models,
as summarized in Fig.~\ref{outgap}.
However, in examining only the presence of peaks at
the observed phases, we ignore the fact that most of these models
produce too many peaks.
For instance, the inclusion of both outward- and inward-going emission
produces four peaks for most geometries.
In addition, including many values of $w$ in the calculations
(which results in many geometries compatible with the observations)
would diffuse the peaks, without fine-tuning the models by
restricting the emission region in azimuth or radius.
The overly liberal approach we take in our modeling, which produces
as many peaks as possible, is intended
only to determine the ability to construct a model which produces
peaks at the right positions, with the assumption that the remaining
degrees of freedom could be used to fine-tune the model to better
fit the entire light curve observed.

One conclusion that is not well represented
in Fig.~\ref{outgap} is that values of $w > 0.40$ do
not contribute significantly to the observed optical
peaks.
\citet{zhang_c1997} predict that because
\psr\ rotates slowly, 70\% of the volume of the outer
magnetosphere should be filled by the outer gap.
While this fractional volume is not directly comparable to
$w$, taken simplistically, it seems to imply that
the optical emission is being emitted well inside (\ie not filling)
the boundaries of the outer gap.

The Crab Pulsar is the only other pulsar for which optical 
polarization measurements are available.
The polarization pulse profile measured in \psr\ is
different in character than that measured in the Crab.
The Crab's linearly polarized {\it flux} is maximized at the
peaks and minimized in the bridge, while the
polarized {\it fraction} is maximized in the bridge and minimized at
the peaks \citep{smith_jdp1988}.
In \psr, the linearly polarized flux and polarized fraction are
both maximized in the bridge and both minimized at the peaks.
The low temporal resolution of our optical measurements will
lead to a decrease in the measured linearly polarized flux,
but our observations of the Crab at the same temporal resolution
show a much greater polarized flux at the peaks (where the
position angle swings rapidly) than in the bridge.
Under the RVM, this decrease in measured linearly polarized
flux would be small, as can be seen by noting that the
position angle sweep is not rapid at phases 0.2 and 0.8 in
Fig.~\ref{rvm}.
However, the RVM does not apply to the position angles measured
in the Crab (as the position angles execute a ``double sweep''), 
so without a more detailed polarization model for
\psr, we cannot rule out the chance that rapid swings have
degraded our measured polarized flux.

The 1-$\sigma$ upper limit on the unpulsed flux is 16\%, limiting
the contribution of thermal radiation to the observed flux.
As we are presenting only differential photometry in these
observations, we must rely on absolute photometry presented by
other authors, which almost certainly have different methods
of background subtraction (and treatment of the nearby extended
object).
As such, we note only that this limit on the thermal radiation
from \psr\ is not surprising, given realistic extrapolations of the
Rayleight-Jeans emission \citep{koptsevich_2001}.
The high pulsed fraction does, however, rule out models in
which the optical emission is due to a fallback disk
(Perna, Hernquist, \& Narayan 2000)\nocite{perna_hn2000}. 
A comparable pulsed / unpulsed measurement in the UV would
provide a much better constraint on the Rayleigh-Jeans tail
of the thermal emission from the surface of \psr.

\acknowledgments

We were greatly assisted by Stephen Kaye and the Palomar engineering
staff.
Thanks to J. Everett and A. Shearer for additional assistance, and
to J. Weisberg for encouraging and constructive comments.
This work was supported by NSF Grants AST-9618880, AST-9819762,
and AST-0096930.
This investigation made use of observations made with the 
NASA/ESA Hubble Space Telescope, obtained from the data archive 
at the Space Telescope Science Institute. 
STScI is operated by the Association of Universities for 
Research in Astronomy, Inc., under NASA contract NAS 5-26555.




\bibliographystyle{astronbdk}
\bibliography{/users/bdk/pcam/papers/bib/psr0656,/users/bdk/pcam/papers/bib/emissionmodels,/users/bdk/pcam/papers/bib/radiopol,psr0656}

\clearpage


\clearpage

\end{document}